\documentclass[fleqn,12pt]{article}
\usepackage{graphicx}
\usepackage{eucal}
\usepackage{amsfonts}
\usepackage{amssymb}
\usepackage{amsbsy}
\usepackage{amsmath}
\usepackage[numbers,compress]{natbib}
\usepackage{mathpazo}
\usepackage{color}
\def\re#1{(\ref{#1})}

\begin{document}

\title{\bf Microinertia and internal variables}
\author{Arkadi Berezovski$^1$ and Peter V\'an$^{234}$\\
\small $^1$ Centre for Nonlinear Studies\\
\small Institute of Cybernetics at Tallinn
\small University of Technology\\
\small Akadeemia tee 21, 12618 Tallinn, Estonia\\
\small $^2$ Wigner Research Centre for Physics,
\small Institute of Particle and Nuclear Physics\\
\small Department of Theoretical Physics,\\
\small H-1525, Budapest, P.O. Box 49, Hungary\\
\small $^3$ BME, Department of Energy Engineering\\
\small H-1521 Budapest, P.O. Box 91, Hungary\\
\small $^4$ Montavid Thermodynamic Research Group
}

\date{\today}
\maketitle
\begin{abstract}
The origin of microinertia of micromorphic theories is investigated from the 
point of view of non-equilibrium thermodynamics. In the framework of  dual 
internal variables microinertia stems from a thermodynamic equation of state 
related to the internal variable with the properties of mechanical momentum.
\end{abstract}

\newcommand{\mvec}[1]{\mbox{\boldmath{$#1$}}}

\section{Introduction}

The prediction of a thermomechanical behavior of solids with microstructure cannot be completely correct unless the influence of the microstructure is taken into account.
In spite of the possible small scale characterizing the microstructure, its
influence may be not necessary small if corresponding gradients are comparable
with those at the macroscale.
It is well understood that microstructural effects can be described only by
means of a generalization of classical thermomechanical theory.
Such a generalization is usually achieved at the expense of the extension of
the state space by additional independent variables like the microrotation in
Cosserat's media, the microdeformation in micromorphic media, or the heat flux
in extended thermodynamics.

Making the distinction between the medium and its microstructural components,
one implicitly admits that they can respond to dynamic loads differently.
This is reflected in the notion of microinertia appeared in the description of
micromorphic media \cite{eringen64, mindlin64}.
Mindlin \cite{mindlin64} described the deformation of what he called "cell"
explicitly, while Eringen and Suhubi \cite{eringen64} introduced an averaging
procedure over the microelement.
Both approaches results in similar balances of micromomentum in the linear case.
As noted by Eringen and Suhubi \cite{eringen64}, the corresponding microinertia tensor "resembles" the physical inertia.
This means that the coefficient of the second time derivative of the
microdeformation is consistent with the constitutive relations.

Mindlin as well as Eringen and Suhubi considered their microdeformation as an
additional degree of freedom.
Therefore, it should be governed (controlled) by boundary conditions pointed
out in the general form by both approaches.
However, it is difficult to imagine boundary conditions that excite the
microstructure only, without affecting the medium itself.
This difficulty brings the internal degrees of freedom together with internal variables of state, which even cannot be controlled by boundary conditions.

The classical treatment of internal variables suggests that their evolution
equations may contain only the first-order time derivatives, and, therefore,
have no relations to inertia \cite{coleman67, kestin93, maugin94, maugin06}.
This is definitely true for order parameters considered as internal variables,
{and for pure dissipative effects like heat conduction.}
However, the generalization of the internal variables theory by the dual
internal variables concept allows hyperbolic evolution equations as well
\cite{van08}.
The question is how the coefficient of the second order time derivative of the
microdeformation corresponds to microinertia.

In order to answer this question, we will show how the formalism of dual
internal variables works.
Starting with simplest possible case, we demonstrate the difference between the standard single internal variable theory and the dual internal variables concept.
Then we compare the generalization of linear elasticity by dual internal variables with the Mindlin microelasticity model \cite{mindlin64}.
For simplicity, all the considerations are made in one-dimensional case.

\section{Single internal variable: one-dimensional example}

While for mechanical motion of a material point it is sufficient to prescribe
its coordinates and velocities, the theory of internal variables is
essentially thermodynamical.
This means that we deal with a thermodynamical system and suppose that all
thermodynamic quantities like temperature, energy, entropy,
etc. are defined.

\subsection{ Entropic representation}
Here we assume that the entropy density $S$ is
specified as a function of the internal energy density $E$  and an internal
variable $\varphi$ and its space derivative $\varphi_x$:
\begin{equation}\label{4}
S=\overline{S}(E,\varphi,\varphi_{x}).
\end{equation}
Constitutive assumption (\ref{4}) allows us to introduce the corresponding
conjugate quantities as follows:
\begin{equation}\label{5}
\frac{1}{\theta}:=\frac{\partial \overline{S}}{\partial E}, \quad
\frac{\tau}{\theta}:=\frac{\partial \overline{S}}{\partial \varphi}, \quad
\frac{\eta}{\theta}:=\frac{\partial \overline{S}}{\partial \varphi_{x}},
\end{equation}
where $\theta$  is the temperature.
With the usual thermodynamic notation with differentials the consequent Gibbs relation can be written as
\begin{equation}\label{51}
d E = \theta d S -\tau d\varphi -\eta d\varphi_x.
\end{equation}
Therefore, Eq. \re{4} is the equation of state of a "gradient" (or weakly
non-local) internal variable theory.

The conservation  of the internal energy can be represented as
\begin{equation}\label{6}
E_{t}+Q_{x}=0,
\end{equation}
where  ${Q}$ is the 
heat flux, index $t$ denotes time derivative.
For the calculation of the entropy production a generalized entropy current
density (i.e. entropy flux) is introduced in the form
\begin{equation}\label{ecmod1}
J = \frac{Q-\eta\varphi}{\theta}.
\end{equation}
Here the deviation of the entropy current density from its classical form
$J_{class} = Q/\theta$ is represented by a particular additional term.
This form is characteristic for second-order weakly nonlocal
theories and can be justified either by a rigorous exploitation of the entropy
inequality \cite{van05,VanPap10a}, or by the classical separation of
divergences \cite{GroMaz62,maugin06}. 
In the next section we will demonstrate the second method.

With the help of  modified entropy flux \re{ecmod1}
the entropy production can be calculated as follows:
\begin{equation}\label{sbal}
\begin{split}
  S_{t}+J_{x} & =\frac{1}{\theta}E_t + \frac{\tau}{\theta}\varphi_t +
\frac{\eta}{\theta}\varphi_{tx} + \left(\frac{Q-\eta\varphi}{\theta}\right)_x \\
    & =Q \left(\frac{1}{\theta}\right)_x +
	\varphi_t \left(\frac{\tau}{\theta}
	-\left[\frac{\eta}{\theta}\right]_x\right) \\
    & =\left(Q- \eta\varphi_t \right)\left(\frac{1}{\theta}\right)_x +
	\frac{\varphi_t}{\theta} \left(\tau -\eta_x\right) \geq 0.
\end{split}
%
\end{equation}
A constitutive relation  for heat flux and an evolution equation for internal variable are needed to be specified for any considered process.
The given form of the entropy production supposes two possible representations
of the thermodynamic fluxes and forces as one can see inspecting the last two
lines of Eq. \re{sbal}.
In the first form the thermal part is the classical, while in the second form
the thermal part is modified. 
However, the linear constitutive equations are identical, as it is shown in the Appendix.

\subsection{Energetic representation}

Free energy is the Legendre transformation of the internal energy by the
entropy $W=E-S\theta$.
It is convenient for practical use, because temperature
is the introduced variable of state instead of  entropy or internal energy:
\begin{equation}\label{41}
W=\overline{W}(\theta,\varphi,\varphi_{x}).
\end{equation}
Partial derivatives of $W$ define
\begin{equation}\label{5f}
S:=-\frac{\partial \overline{W}}{\partial \theta}, \quad
\tau:=-\frac{\partial \overline{W}}{\partial \varphi}, \quad
{\eta}:=-\frac{\partial \overline{W}}{\partial \varphi_{x}},
\end{equation}
which results in the Pfaffian form
\begin{equation}\label{52}
d W = -Sd\theta -\tau d\varphi -\eta d\varphi_x.
\end{equation}
The energy conservation equation is
\begin{equation}\label{7}
(S\theta)_{t}+Q_{x}=h^{int}, \quad h^{int}:=-W_{t},
\end{equation}
where the right-hand side of Eq. (\ref{7})$_1$ is formally  an {internal  heat source} \cite{maugin06}.

The energy conservation equation should be accompanied by the {second law of
thermodynamics}, here written as
\begin{equation}\label{8}
S_{t} + ({Q}/\theta + { K})_{x}\geq 0,
\end{equation}
where in the entropy flux  is additively separated onto the classical part and
the extra entropy flux $K$.
Multiplying the second law (\ref{8}) by $\theta$
\begin{equation}\label{9}
\theta S_{t} + \theta ({Q}/\theta)+{ K})_{x}\geq 0,
\end{equation}
and taking into account Eq. (\ref{7}), we obtain
\begin{equation}\label{10}
-\left(W_{t}+S\theta_{ t}\right)+
(\theta {K})_{x} -({Q}/\theta)+{ K})\theta_{x} \geq 0.
\end{equation}
The last equation can be represented in the form
\begin{equation}\label{11}
S{\theta}_{t} + ({Q}/\theta)+{ K})\theta_{x} \leq h^{int}+(\theta {K})_{x}.
\end{equation}
The internal heat source $h^{int}$ is calculated  following the constitutive
assumption (\ref{41})
\begin{equation}\label{12}
h^{int}=- W_{t}=
-\frac{\partial W}{\partial \theta}\theta_{t}
-\frac{\partial W}{\partial \varphi}\varphi_{t}
-\frac{\partial W}{\partial \varphi_{x}}\varphi_{xt}
=S{\theta}_{t}+\tau{\varphi}_{t}+\eta {\varphi}_{xt}=h^{th}+h^{intr}.
\end{equation}
Accounting for Eq. (\ref{12}), dissipation inequality (\ref{11}) can be rewritten as
\begin{equation}\label{13}
\Phi= \tau{\varphi}_{t}+\eta{\varphi}_{xt}-(Q/\theta+K)\theta_{x}
+(\theta {K})_{x}\geq0.
\end{equation}
Here $\Phi$ is the entropy production multiplied by the temperature.
To rearrange the dissipation inequality, we add and subtract the same term
$\eta_{x}{\varphi}_{t}$
\begin{equation}\label{14}
\Phi= \tau{\varphi}_{t}+\eta\dot{\varphi}_{x}
-\eta_{x}{\varphi}_{t}+\eta_{x}{\varphi}_{t}
-(Q/\theta+K)\theta_{x}
+(\theta {K})_{x}\geq0,
\end{equation}
that leads to
\begin{equation}\label{15}
\Phi= (\tau-\eta_{x}){\varphi}_{t}
-(Q/\theta+K)\theta_{x}
+(\eta{\varphi}_{t}+\theta {K})_{x}\geq0.
\end{equation}
Therefore we can define the "extra" entropy flux by the elimination of the
divergence term in Eq. (\ref{15})
\begin{equation}\label{16}
{K}=-\theta^{-1}\eta{\varphi}_{t}.
\end{equation}
This method, the separation of divergences, has its roots in the classical
irreversible thermodynamics \cite{GroMaz62} and formulated explicitly in 
case of internal variables by Maugin \cite{maugin90}.
Then the dissipation inequality reduces to
\begin{equation}\label{17}
\Phi= (\tau-\eta_{x}){\varphi}_{t}
-(Q+\eta{\varphi}_{t})(\log(\theta))_{x}
\geq0.
\end{equation}
It is remarkable that in the isothermal case ($\theta_{x}=0$) the dissipation
is determined by the internal variable only. 
It is easy to check that inequalities \re{17} and \re{sbal} are identical.

\subsection{Evolution equation for a single internal variable}

In the \emph{isothermal} case  the dissipation inequality (\ref{17}) is
\begin{equation}\label{18}
\Phi=(\tau-\eta_{x}){\varphi}_{t}\geq0,
\end{equation}
The linear solution is 
\begin{equation}\label{19}
{\varphi}_{t}=k (\tau-\eta_{x}),  \quad k \geq0,
\end{equation}
since  dissipation inequality  (\ref{18})  is satisfied automatically in this case
\begin{equation}\label{20}
\Phi=k{\varphi}_{t}^{2}\geq0, \quad \mbox{if} \quad  k \geq0.
\end{equation}
It is easy to see that the dissipation is the product of the thermodynamic flux ${\varphi}_{t}$ and the thermodynamic force $(\tau-\eta_{x})$.
The proportionality between the thermodynamic flux and the conjugated force is the standard choice to satisfy the dissipation inequality.

To see how the obtained evolution equation looks like, we specialize free energy dependence (\ref{4}) in the isothermal case to a quadratic one
\begin{equation}\label{21}
      \overline{W} =  \frac{1}{2} B \varphi^{2} + \frac{1}{2} C \varphi^{2}_{x},
\end{equation}
where $B$ and $C$ are material parameters.
It follows from  equations of state (\ref{5}) that
\begin{equation}\label{22}
      \tau:=-\frac{\partial \overline{W}}{\partial \varphi}=-B \varphi, \quad
{\eta}:=-\frac{\partial \overline{W}}{\partial \varphi_{x}}=-C \varphi_{x},
\end{equation}
and evolution equation (\ref{19})  is an equation of reaction-diffusion type
\begin{equation}\label{23}
{\varphi}_{t}=k (C\varphi_{xx}-B \varphi),  \quad k \geq0.
\end{equation}
The given standard  formalism  of the single internal variable of state is
sufficient for many cases \cite{rice71, lubliner73, sidoroff88, muschik94, houlsby00,
rahouadj03, lebon08, horstemeyer10}. 
{ In the case of a more general free
energy,  Eq. \re{23} is the Ginzburg-Landau equation \cite{Van02}}.

\subsection{Dual internal variables in one dimension}

Now we would like to extend the same technique to the case of two internal
variables. 
However, none of them is presupposed to be definitely an internal
variable of state or an internal degree of freedom. 
Let us consider them as internal variables "in general".  
Here we treat the isothermal case, therefore we extend the free energy focused  
train of thought of the previous section.
The corresponding entropic treatment is straightforward and equivalent.

Let us suppose that the free energy density depends on the internal variables
$\varphi, \psi$ and their space derivatives
\begin{equation}\label{eq75}
W=\overline{W}(\theta,\varphi,\varphi_{x},\psi,\psi_{x}).
\end{equation}
The  equations of state in the case of two internal variables read
\begin{equation}\label{eq76}
S=-\frac{\partial \overline{W}}{\partial \theta}, \quad
\tau:=-\frac{\partial \overline{W}}{\partial \varphi}, \quad
{\eta}:=-\frac{\partial \overline{W}}{\partial \varphi_{x}}, \quad
\xi:=-\frac{\partial \overline{W}}{\partial \psi}, \quad
\zeta:=-\frac{\partial \overline{W}}{\partial \psi_{x}}.
\end{equation}
The Gibbs relation follows accordingly
\begin{equation}\label{53}
d W = -Sd\theta -\tau d\varphi -\eta d\varphi_x - \xi d \psi- \zeta d\psi_x.
\end{equation}
Then we consider a non-zero extra entropy flux following the case of a single
internal variable and set
\begin{equation}\label{eq103}
{K}=-\theta^{-1}\eta{\varphi}_{t}-\theta^{-1}\zeta{\xi}_{t}.
\end{equation}
It can be checked that  the intrinsic heat source
is determined in the considered case as follows
\begin{equation}\label{eq1009}
 \widetilde{h}^{intr}:=(\tau-\eta_{x}){\varphi}_{t}+(\xi-\zeta_{x}){\psi}_{t}.
\end{equation}
The latter means that the dissipation inequality in the isothermal
case reduces to
\begin{equation}\label{eq10090}
\Phi=(\tau-\eta_{x}){\varphi}_{t}+(\xi-\zeta_{x}){\psi}_{t}\geq0.
\end{equation}
{
The solution of the dissipation inequality can be represented as \cite{van08}
\begin{equation}\label{eq111}
\begin{pmatrix}
 {\varphi}_{t} \\
 {\psi}_{t} \\
\end{pmatrix}
= \mathbf{L}
\begin{pmatrix}
  \tau-\eta_{x} \\
  \xi-\zeta_{x} \\
\end{pmatrix},
\quad \mbox{where} \quad
\mathbf{L}
=
\begin{pmatrix}
  L_{11} & L_{12} \\
  L_{21} & L_{22} \\
\end{pmatrix}.
%
\end{equation}
Nonnegativity of the entropy production (\ref{eq10090}) results in the
positive semidefiniteness of the symmetric part of the
conductivity matrix $\mathbf{L}$, which requires
\begin{equation}\label{n1}
    L_{11}\geq0,\quad  L_{22}\geq0,  \quad L_{11} L_{22} -
    \frac{(L_{12}+L_{21})^2}{2}\geq0.
\end{equation}
To be more specific, we keep a quadratic free energy density
\begin{equation}\label{eq77}
      \overline{W} = \frac{1}{2} B \varphi^{2} + \frac{1}{2} C \varphi^{2}_{x}+\frac{1}{2} D
      \psi^{2}+\frac{1}{2} F  \psi^{2}_{x}.
\end{equation}
Calculating the quantities defined in Eq. \re{eq76}
\begin{equation}\label{eq76+}
\tau:=-\frac{\partial \overline{W}}{\partial \varphi} = -B \varphi, \quad
{\eta}:=-\frac{\partial \overline{W}}{\partial \varphi_{x}}= -C \varphi_{x},
\end{equation}
\begin{equation}\label{eq76++}
\xi:=-\frac{\partial \overline{W}}{\partial \psi}=-D
      \psi, \quad
\zeta:=-\frac{\partial \overline{W}}{\partial \psi_{x}}=-F  \psi_{x},
\end{equation}
we can represent system of Eqs. (\ref{eq111}) in the form
\begin{eqnarray}\label{int1}
{\varphi}_{t} &=& L_{11}(-B \varphi+C \varphi_{xx}) + L_{12}(-D \psi+F
\psi_{xx}), \\ \label{int2}
{\psi}_{t} & =& L_{21}(-B \varphi+C \varphi_{xx}) + L_{22}(-D \psi+F
\psi_{xx}).
\end{eqnarray}
Now we will derive a single evolution equation for the primary internal variable. 
To this end a suitable rearrangement of the previous equations is performed:
\begin{eqnarray}\label{rint1}
 \left(\frac{\partial}{\partial t} +
	 L_{11}(B - C \frac{\partial^2}{\partial x^2})\right)\varphi +
	 L_{12}\left(D - F \frac{\partial^2}{\partial x^2}\right)\psi  &=&
	 \widehat L_{11} \varphi + \widehat L_{12} \psi = 0, \\ \label{rint2}
 L_{21}\left(B - C \frac{\partial^2}{\partial x^2}\right)\varphi +
	 \left(\frac{\partial}{\partial t} + L_{22}(D -F\frac{\partial^2}{\partial
	 x^2})\right)\psi  &=&
	 \widehat L_{21} \varphi + \widehat L_{22} \psi = 0.
\end{eqnarray}
Here the traditional notation of time and space derivatives is used
for the clear distinction of the differential operators  $\widehat L_{11}$, 
$\widehat
L_{12}$, $\widehat L_{21}$ and $\widehat L_{22}$. 
Then the easiest way to eliminate $\psi$ is to multiply the first equation, 
\re{rint1}, by $\widehat L_{22}$ and the second one, \re{rint2},  by $\widehat 
L_{12}$ and subtract them:
\begin{equation}\label{feq}
\begin{split}
  \widehat L_{22}\widehat L_{11}\varphi - \widehat L_{12}\widehat L_{21}\varphi 
  & ={\varphi}_{tt} - \det L(CD + BF)\varphi_{xx}+\\
   +(BL_{11}+D L_{22}){\varphi}_{t} &-
(CL_{11}+F L_{22}){\varphi}_{txx} +\det L  \left(BD \varphi   + CF\varphi_{xxxx}\right)=0,
\end{split}
\end{equation}
where $\det L = L_{11}L_{22}-L_{12}L_{21}$ is the determinant of $L$. 
The free energy
density $W$ is non-negative by default, which results in non-negativity of
material parameters $B, C, D$, and $F$.
This means that Eq. (\ref{feq}) is the hyperbolic wave equation with dispersion
and dissipation.

Thus, extending the state space of our thermodynamic system by an additional internal variable and keeping the quadratic form for the free energy density, we arrive at the hyperbolic evolution equation for the primary internal variable.
The corresponding evolution equation for the secondary internal variable can be derived similarly.
The natural question is the following: can we associate the abstract internal variables with the description of any physical process?

\subsection{Example I: Linear elasticity}

To answer the question posed above, we simplify the situation as much as possible.
We note first that dissipation will be absent if the diagonal components of the 
conductivity matrix are zero ($L_{11} = L_{22}=0$).
Due to  inequalities \re{n1} this requires antisymmetric conductivity
matrix, that is, Casimir type reciprocal relations $L_{12}= -L_{21}$.
Next, dispersion will be eliminated if we choose the material parameters $B=0$ and
$F=0$.


It follows immediately  that  the evolution equations for the dual internal variables  (\ref{int1}) and (\ref{int2}) can be rewritten as
\begin{equation}\label{eq115a-}
{\varphi}_{t}=-L_{12}D \psi,
\end{equation}
\begin{equation}\label{eq117a-}
{\psi}_{t}=-L_{12}C(\varphi_{x})_{x}. 
\end{equation}
Without loss of generality, we can choose the value $L_{12}=-1$.
We should emphasize that the two evolution equations express the duality between internal variables: one internal variable is driven by another one and vice versa.

Now we are ready to relate the introduced internal variables to  well-known
physical quantities.
First, we note that Eqs. (\ref{eq115a-}), (\ref{eq117a-})  can be represented
in the form
\begin{equation}\label{eq115a+-}
{\varphi_{xt}}=D \psi_{x},
\end{equation}
\begin{equation}\label{eq117a+-}
{\psi}_{t}=C(\varphi_{x})_{x}. 
\end{equation}
Secondly, denoting $\varphi_{x}$ as $\varepsilon$, and $\psi$ as $\rho v$ we
arrive at the system of equations
\begin{equation}\label{intel1}
{\varepsilon}_{t}=D (\rho v)_{x},
\end{equation}
\begin{equation}\label{intel2}
(\rho v)_{t}=C\varepsilon_{x}, 
\end{equation}
which  coincides with one-dimensional equations for linear elasticity (with the
standard notation for derivatives), and $\rho$ is constant:
\begin{equation}\label{eq85}
        \frac{\partial \varepsilon}{\partial t} =\frac{\partial v}{\partial x},
\end{equation}
\begin{equation}\label{eq86}
        \rho \frac{\partial v}{\partial t} =\frac{\partial \sigma}{\partial
        x},
        \end{equation}
after the following choice of coefficients: $D=1/\rho$ and $C=\rho c^{2}$, where $c$
is the elastic wave velocity and $\sigma = \rho c^{2} \varepsilon$.
This means that the internal variable $\varphi$ can be interpreted as the displacement and the internal variable $\psi$ as the momentum.
These variables are controlled by boundary conditions and, therefore, they are
true mechanical degrees of freedom, while appeared as internal degrees of
freedom first. 
No doubts arise about inertia in this case.

We have considered a thermodynamic system (body) under assumption that there may be certain internal phenomena which we tried to describe by internal variables.
We have derived possible evolution equations for these internal variables from the first and second laws of thermodynamics.
It has been demonstrated that  evolution equations for internal variables may be hyperbolic for a quadratic free energy density.
Moreover, in the non-dissipative and non-dispersive case these evolution
equations can be identified with equations of linear elasticity.

It follows  that both internal variables of state and internal degrees of freedom can be  considered
as particular cases in the framework of unified formalism.
This formalism can help us to consider generalized continuum models by means of internal variables.
In our next example, we look at the Mindlin micromorphic elasticity theory.

\section{Example II: One-dimensional microelasticity}

Following \cite{mindlin64}, in our one-dimensional treatment we introduce  the
macrostrain $\varepsilon$, the microdeformation $\psi$, the relative
deformation $\gamma=\varepsilon - \psi$, and the microdeformation gradient
$\psi_{x}$.
The corresponding quadratic free energy density  \cite{mindlin64} is reduced to
\begin{equation}\label{mind1}
  W(\varepsilon, \gamma, \psi_{x}) = \frac{1}{2}\rho 
  c^{2}\varepsilon^{2}+g\varepsilon\gamma+\frac{b}{2}\gamma^{2}+\frac{a}{2} 
  \psi_{x}^{2}.
\end{equation}
Equations of state  determine the corresponding stresses
\begin{equation}\label{mind2}
  \tau = \frac{\partial W}{\partial \varepsilon}=\rho c^{2}\varepsilon+g\gamma, \quad \sigma = \frac{\partial W}{\partial \gamma}= g\varepsilon+b\gamma, \quad  \mu =\frac{\partial W}{\partial \psi_{x}}=a\psi_{x},
\end{equation}
and equations of motion have the form \cite{mindlin64} (without body forces)
\begin{equation}\label{mind3}
  \rho u_{tt}= \tau_{x}+\sigma_{x},
\end{equation}
\begin{equation}\label{mind4}
 \frac{1}{3}\rho'd^{2} \psi_{tt} = \mu_{x}+\sigma,
\end{equation}
where $u$ is the displacement, and $\rho'd^{2}/3$ plays the role of
microinertia.

Now we  compare the Mindlin microelasticity theory with the dual internal
variables approach.
To avoid the confusion with the original notation of Mindlin, we use other
symbols for internal variables.
We suppose that the free energy density depends on the gradient of the
displacement, $u_x$, the internal variables $\alpha, \beta$ and their space
derivatives
\begin{equation}\label{eq75+}
W=\overline{W}(u_{x},\alpha,\alpha_{x},\beta,\beta_{x}).
\end{equation}
Then the equations of state follow
\begin{equation}
\sigma':=\frac{\partial \overline{W}}{\partial u_{x}}, \quad
\tau':=-\frac{\partial \overline{W}}{\partial \alpha}, \quad
{\eta}:=-\frac{\partial \overline{W}}{\partial \alpha_{x}}, \quad
\xi:=-\frac{\partial \overline{W}}{\partial \beta}, \quad
\zeta:=-\frac{\partial \overline{W}}{\partial \beta_{x}}.
\end{equation}
As in \cite{mindlin64}, we use   a quadratic function for the free energy density
\begin{equation}\label{eq78}
      \overline{W} = \frac{1}{2}\rho c^{2} u_{x}^{2} + A \alpha u_{x} + \frac{1}{2} B \alpha^{2} + \frac{1}{2} C \alpha^{2}_{x}+\frac{1}{2} D \beta^{2}.
\end{equation}
Considering the non-dissipative case, we have for the evolution equation for internal variables  (cf. Eqs. (\ref{int1}), (\ref{int2}) with $L_{11} = L_{22}= 0$)
\begin{equation}\label{eq115a}
{\alpha}_{t}=L_{12} (\xi-\zeta_{x})= - L_{12} D\beta,
\end{equation}
\begin{equation}\label{eq117a}
{\beta}_{t}=-L_{12}(\tau'-\eta_{x})= -L_{12}(C \alpha_{xx}-A u_{x} - B \alpha).
\end{equation}
 As a result, we arrive at the hyperbolic evolution equation equation for the primary internal variable
\begin{equation}\label{int}
   I \alpha_{tt}=C \alpha_{xx}-A u_{x} - B \alpha,
\end{equation}
where $ I = 1/(L_{12}^2 D)$ is the measure of microinertia.

The balance of  linear momentum
\begin{equation}\label{eq71}
{\rho} v_{t} = \sigma'_{x},
\end{equation}
results in the form
\begin{equation}\label{eq83}
                      \rho u_{tt}=\rho c^{2} u_{xx} + A \alpha_{x}.
\end{equation}
Let us rewrite the governing equations of the Mindlin microelasticity (\ref{mind3}) and (\ref{mind4}) using the notation $u_{x}$ instead of $\varepsilon$ and $\alpha$ instead of the microdeformation $\psi$. We have, respectively,
\begin{equation}\label{mind5}
  \rho u_{tt}= \tau_{x}+\sigma_{x}= \rho c^{2}u_{xx}+(b+2g)u_{xx}-(g+b)\alpha_{x},
\end{equation}
\begin{equation}\label{mind6}
 \frac{1}{3}\rho'd^{2} \alpha_{tt} = a\alpha_{xx}+(g+b)u_{x}-b\alpha,
\end{equation}
It is easy to see that systems of equations (\ref{int}), (\ref{eq83}) and
(\ref{mind5}), (\ref{mind6}) coincide if the following values of parameters are
chosen: $A=-(b+g), B=b, C=a, I=1/3\rho'd^{2}, b+2g=0$.
This means that the dual internal variable theory can recover the generalized 
continuum theory of Mindlin \cite{mindlin64} and the corresponding microinertia 
measure has the same justification as in the Mindlin microelasticity.

\section{Summary and discussion}

In this short note we have analyzed the relation of mechanical and
thermodynamical approaches of extending the state space of continuum theories.
Our analysis focused on the origin of inertial terms in the evolution equations.
In the first part of the paper we have presented the thermodynamic background
of single and dual internal variable theories both in the entropic  and  in the
energetic representation \cite{callen85}.

The recent methodology of dual internal variables unifies these treatments and
we have shown that both classical linear elasticity and Mindlin type
microelasticity are well suited in this general background without a direct
reference to mechanical principles or notions.
Our treatment was restricted to one spatial dimension and without a
mention of ispace-time related conceptual questions.
The 3D tensorial representation of the application of the dual internal variables approach is given in \cite{bem11gen,bem11temp}.

What is the origin of inertia in the light of our approach?
The two ingredients were the thermodynamic equation of state and the
antisymmetric part of the constitutively constructed dynamics.
The more surprising part is the first: in our treatment inertia and
microinertia stem from an equation of state.
The coefficients $C$ and $D$ in the quadratic free energy function of
linear elasticity,   (Eq. \re{eq77}), are connected with the density, which is
the measure of inertia. 
In case of microinertia the coefficient $D$ is enough,
the gradient term of the second variable was not necessary.
{In our dual internal variable approach one of the variables plays
the role of momentum when compared to non-dissipative mechanics, on the other
hand it is a state variable from a thermodynamic point of view.

This kind of close connection of mechanics and thermodynamics is not too
surprising from the broader perspective of relativistic theories.
Therein momentum cannot be separated from energy in a covariant treatment,
therefore an energy dependent entropy depends naturally on the momentum as well.
This is the root of the famous problem of the proper Lorentz transformation of
the temperature. 
The first formulation of Planck and Einstein considered energy
and momentum as variables of the entropy \cite{Pla07a,Ein07a,Pla08a}. 
Then Blanu\v{s}a and Ott seemingly have eliminated this dependence by fixing the
thermometer to the body \cite{Bla47a,Ott63a}. 
However, in a proper covariant treatment these approaches are only two sides of 
the same coin, where the entropy depends on energy-momentum 
\cite{BirVan10a,VanBir12a}.

In our nonrelativistic treatment the integration of mechanics and
thermodynamics looks like more involved than usually taken.}
From the point of view of the presented approach it is remarkable, that Hamiltonian dynamics is a particular case characterized by zero dissipation.
{Moreover, the possibility of a pure variational description} can be
a consequence of ideal dynamics without dissipation \cite{van09}.

In summary, we have shown that inertial terms are natural in a
thermodynamic theory with dual variables and the conditions of their appearance
is well understandable in terms of mechanical notions.
However, a true space-time related frame independent treatment is required to a final clarification of this issue.

\section*{Acknowledgments}

The work was supported by the EU through the European Regional Development Fund, by the Estonian Research Council grant  PUT434 (A.B.), by the grant OTKA K104260 (P.V.), and by the Estonian-Hungarian Academic Joint Research Project SNK-66/2013 (A.B. and P.V.).

\section{Appendix: { Representations} of entropy production}

{ In some cases entropy production (\ref{sbal})$_2$  appears} in
the following form
\begin{equation}\label{ep1}
\Sigma = J_1 X_1 + J_2 (X_2 - A X_1) \geq 0,
\end{equation}
where $J_1$ and $J_2$ are constitutive quantities, and they are to be
determined as functions of $X_1$ and $X_2$,
$A$ is a given function of $X_1$ and $X_2$.
One may introduce the following choice of thermodynamic fluxes and
forces:
\begin{center}
\begin{tabular}{c|c|c}
Fluxes & $J_1$ &  $J_2$ \\ \hline
Forces & $X_1$ &  $X_2- A X_1$ \\
\end{tabular}\\\vskip .21cm
{Table 1. First choice of thermodynamic fluxes and forces}\end{center}
The linear solution of inequality (\ref{ep1}) results in:
\begin{eqnarray}
J_1 &=& l_1 X_1 +l_{12} (X_2-A X_1), \label{l1}\\
J_2 &=& l_{21} X_1 +l_{2} (X_2-A X_1). \label{l2}
\end{eqnarray}
Here the inequality requires that the symmetric part of the coefficient matrix
is positive definite, therefore $l_1, l_2, l_{12}$ and $l_{21}$
coefficients fulfill the following inequalities:
\begin{equation}\label{ine1}
l_1>0, \quad l_2>0, \quad \text{and} \quad
l_1 l_2 - \frac{(l_{12}+l_{21})^2}{4} \geq 0.
\end{equation}
However, the entropy production may be written in a  distinct but equivalent
form
\begin{equation}\label{ep2}
\Sigma = (J_1 - A J_2)X_1 + J_2 X_2 \geq 0.
\end{equation}
This grouping of the terms suggest a different choice of the thermodynamic
forces and fluxes:
\begin{center}
\begin{tabular}{c|c|c}
Fluxes & $J_1 -A J_2$ &  $J_2$ \\ \hline
Forces & $X_1$ &  $X_2$ \\
\end{tabular}\\\vskip .21cm
{Table 1. Second choice of thermodynamic fluxes and forces}\end{center}
The consequent linear solution introduces the constitutive relations as follows:
\begin{eqnarray}
J_1 - A J_2 &=& k_1 X_1 +k_{12} X_2, \label{k1}\\
J_2 &=& k_{21} X_1 +k_{2} X_2, \label{k2}
\end{eqnarray}
where the coefficient matrix must fulfill the following inequalities:
\begin{equation}\label{ine2}
k_1>0, \quad k_2>0, \quad \text{and} \quad
k_1 k_2 - \frac{(k_{12}+k_{21})^2}{4} \geq 0.
\end{equation}
The material coefficients $l_1, l_2, l_{12}$ and $l_{21}$ are not independent
of $k_1, k_2, k_{12}$ and $k_{21}$ and their relations can be expressed easily:
\begin{equation}\label{m1}
\begin{pmatrix}
l_1 & l_{12} \\ l_{21} & l_2
\end{pmatrix} =
\begin{pmatrix}
k_1 + A(k_{12}+k_{21}) + A^2 k_2 & k_{12} +A k_2 \\
k_{21} +A k_2 & k_2
\end{pmatrix}
\end{equation}
The inverted form is also explicit
\begin{equation}\label{m2}
\begin{pmatrix}
k_1 & k_{12} \\ k_{21} & k_2
\end{pmatrix} =
\begin{pmatrix}
l_1 - A(l_{12}+l_{21}) + A^2 l_2 & l_{12} -A l_2 \\
l_{21} -A l_2 & l_2
\end{pmatrix}
\end{equation}
As one can check, inequalities \re{ine1} and \re{ine2} are equivalent
and therefore the constitutive equations \re{l1}-\re{l2} are equivalent to
\re{k1}-\re{k2}.
The seemingly different thermodynamic fluxes and forces
lead to equivalent constitutive relations.

\bibliographystyle{unsrt}

\begin{thebibliography}{10}

\bibitem{eringen64}
A~Cemal Eringen and ES~Suhubi.
\newblock Nonlinear theory of simple micro-elastic solids--{I}.
\newblock {\em International Journal of Engineering Science}, 2(2):189--203,
  1964.

\bibitem{mindlin64}
Raymond~David Mindlin.
\newblock Micro-structure in linear elasticity.
\newblock {\em Archive for Rational Mechanics and Analysis}, 16(1):51--78,
  1964.

\bibitem{coleman67}
Bernard~D Coleman and Morton~E Gurtin.
\newblock Thermodynamics with internal state variables.
\newblock {\em The Journal of Chemical Physics}, 47(2):597--613, 1967.

\bibitem{kestin93}
J~Kestin.
\newblock Internal variables in the local-equilibrium approximation.
\newblock {\em Journal of Non-Equilibrium Thermodynamics}, 18:360--379, 1993.

\bibitem{maugin94}
G{\'e}rard~A Maugin and Wolfgang Muschik.
\newblock Thermodynamics with internal variables. {P}art {I}. {G}eneral
  concepts.
\newblock {\em Journal of Non-Equilibrium Thermodynamics}, 19:217--249, 1994.

\bibitem{maugin06}
G{\'e}rard~A Maugin.
\newblock On the thermomechanics of continuous media with diffusion and/or weak
  nonlocality.
\newblock {\em Archive of Applied Mechanics}, 75(10-12):723--738, 2006.

\bibitem{van08}
P.~V\'an, Arkadi Berezovski, and J{\"u}ri Engelbrecht.
\newblock Internal variables and dynamic degrees of freedom.
\newblock {\em Journal of Non-Equilibrium Thermodynamics}, 33(3):235--254,
  2008.

\bibitem{van05}
P.~V{\'a}n.
\newblock Exploiting the second law in weakly non-local continuum physics.
\newblock {\em Mechanical Engineering}, 49(1):79--94, 2005.

\bibitem{VanPap10a}
P.~V{\'a}n and Christina Papenfuss.
\newblock Thermodynamic consistency of third grade finite strain elasticity.
\newblock {\em Proceedings of the Estonian Academy of Sciences},
  59(2):126--132, 2010.

\bibitem{GroMaz62}
Sybren~Ruurds De~Groot and Peter Mazur.
\newblock {\em Non-equilibrium {T}hermodynamics}.
\newblock North-Holland, 1962.

\bibitem{maugin90}
G{\'e}rard~A Maugin.
\newblock Internal variables and dissipative structures.
\newblock {\em Journal of Non-Equilibrium Thermodynamics}, 15(2):173--192,
  1990.

\bibitem{rice71}
James~R Rice.
\newblock Inelastic constitutive relations for solids: an internal-variable
  theory and its application to metal plasticity.
\newblock {\em Journal of the Mechanics and Physics of Solids}, 19(6):433--455,
  1971.

\bibitem{lubliner73}
J~Lubliner.
\newblock On the structure of the rate equations of materials with internal
  variables.
\newblock {\em Acta Mechanica}, 17(1-2):109--119, 1973.

\bibitem{sidoroff88}
F~Sidoroff.
\newblock Internal variables and phenomenological models for metals plasticity.
\newblock {\em Revue de physique appliqu{\'e}e}, 23(4):649--659, 1988.

\bibitem{muschik94}
G{\'e}rard~A Maugin and Wolfgang Muschik.
\newblock Thermodynamics with internal variables. {P}art {II}. {A}pplications.
\newblock {\em Journal of Non-Equilibrium Thermodynamics}, 19:250--289, 1994.

\bibitem{houlsby00}
GT~Houlsby and AM~Puzrin.
\newblock A thermomechanical framework for constitutive models for
  rate-independent dissipative materials.
\newblock {\em International Journal of Plasticity}, 16(9):1017--1047, 2000.

\bibitem{rahouadj03}
Rachid Rahouadj, Jean-Fran{\c{c}}ois Ganghoffer, and Christian Cunat.
\newblock A thermodynamic approach with internal variables using {L}agrange
  formalism. {P}art {I}: General framework.
\newblock {\em Mechanics Research Communications}, 30(2):109--117, 2003.

\bibitem{lebon08}
G~Lebon, D~Jou, and J~Casas-V{\'a}zquez.
\newblock {\em Understanding Non-equilibrium Thermodynamics: Foundations,
  Applications, Frontiers}, chapter 8: Theories with internal variables, pages
  215--236.
\newblock Springer, 2008.

\bibitem{horstemeyer10}
Mark~F Horstemeyer and Douglas~J Bammann.
\newblock Historical review of internal state variable theory for inelasticity.
\newblock {\em International Journal of Plasticity}, 26(9):1310--1334, 2010.

\bibitem{Van02}
P.~V{\'a}n.
\newblock Weakly nonlocal irreversible thermodynamics?the
  {G}inzburg--{L}andau equation.
\newblock {\em Technische Mechanik}, 22(2):104--110, 2002.

\bibitem{callen85}
Herbert~B Callen.
\newblock {\em Thermodynamics and an Introduction to Thermostatistics}.
\newblock Wiley, 1985.

\bibitem{bem11gen}
Arkadi Berezovski, J{\"u}ri Engelbrecht, and G{\'e}rard~A Maugin.
\newblock Generalized thermomechanics with dual internal variables.
\newblock {\em Archive of Applied Mechanics}, 81(2):229--240, 2011.

\bibitem{bem11temp}
Arkadi Berezovski, J{\"u}ri Engelbrecht, and G{\'e}rard~A Maugin.
\newblock Thermoelasticity with dual internal variables.
\newblock {\em Journal of Thermal Stresses}, 34(5-6):413--430, 2011.

\bibitem{Pla07a}
M.~Planck.
\newblock Zur {D}ynamik bewegter {S}ysteme.
\newblock {\em Sitzungsberichte der k\"oniglich Preussische Akademie der
  Wissenschaften}, pages 542--570, 1907.

\bibitem{Ein07a}
A.~Einstein.
\newblock {\"U}ber das {R}elativit\"atsprinzip und die aus demselben gezogenen
  {F}olgerungen.
\newblock {\em Jahrbuch der Radioaktivit\"at und Elektronik}, 4:411--462, 1907.

\bibitem{Pla08a}
M.~Planck.
\newblock Zur {D}ynamik bewegter {S}ysteme.
\newblock {\em Annalen der Physik}, 331(6):1--34, 1908.

\bibitem{Bla47a}
D.~Blanu\v{s}a.
\newblock Sur les paradoxes de la notion d'\'energie.
\newblock {\em Glasnik mat. fiz.; astr.}, 2(4-5):249--50, 1947.

\bibitem{Ott63a}
H.~Ott.
\newblock Lorentz-{T}ransformation der {W}\"arme und der {T}emperatur.
\newblock {\em Zeitschrift f\"ur Physik}, 175:70--104, 1963.

\bibitem{BirVan10a}
T.~S. B\'ir\'o and P.~V\'an.
\newblock About the temperature of moving bodies.
\newblock {\em EPL}, 89:30001, 2010.
\newblock arXiv:0905.1650v1.

\bibitem{VanBir12a}
P.~V\'an and T.S. Bir\'o.
\newblock First order and generic stable relativistic dissipative
  hydrodynamics.
\newblock {\em Physics Letters B}, 709(1-2):106--110, 2012.
\newblock arXiv:1109.0985[nucl-th].

\bibitem{van09}
P.~V{\'a}n.
\newblock Weakly nonlocal non-equilibrium thermodynamics--variational
  principles and {S}econd {L}aw.
\newblock In Tarmo Soomere and Evald Quak, editors, {\em Applied Wave
  Mathematics}, pages 153--186. Springer, 2009.

\end{thebibliography}

\end{document}